\documentstyle [12pt]{article}
\title{
\hspace{3.0truein}{}\\
\vspace{1.0truein}
Critical Statistical Charge for \\Anyonic Superconductivity}
 \author{James Ball\\
          Department of Physics\\
          University of Utah\\
          Salt Lake City,  Utah 84112\\
 and\\
Wei Chen\\
          Department of Physics\\
          University of California\\
Davis, CA 95616}%
\date{}
\begin{document}
\maketitle

\vspace{0.2truein}

\begin{abstract}
We examine a criterion for the anyonic superconductivity at zero 
temperature in Abelian matter-coupled Chern-Simons gauge field 
theories in three dimensions. 
By solving the Dyson-Schwinger equations, 
we obtain a critical value of the statistical charge 
for the superconducting phase in a massless 
fermion-Chern-Simons model. 
\end{abstract}

\newpage 
\baselineskip=24.0truept
One of the remarkable features of 2+1 dimensional gauge theories is that       
they admit a parity (P) and time-reversal (T) violating Chern-Simon (CS) 
term of the gauge field \cite{DJT}. 
Via a CS coupling, the charged planar particles are attached 
with magnetic fluxes and, depending on the strength of the effective
CS coupling, the spin and statistics of the particles 
are transmuted. 
Therefore, coupling a CS field to an ordinary fermion or boson field 
provides a description of anyons \cite{Wil}-\cite{Mat}. 
The concept of anyons has found its application 
first in the fractional quantum Hall effects \cite{PG}-\cite{Wu}. 
It was also conjectured \cite{Lau} that
the gas of charged anyons would exhibit the property of superconductivity. 
In the last few years, much progress has 
been made  along the line \cite{FHL}-\cite{FI}.
In particular, a criterion for the anyonic superconductivity has been
established. In the language of the CS matter field theory, this criterion 
is expressed as \cite{BL}: superconductivity (at zero temperature) occurs 
if and only if the renormalized CS term vanishes. 
This statement is based on the observation that, when the renormalized CS 
term is absent, there exists a Goldstone pole in 
the effective low-momentum lagrangian, and 
the remaining path integral may be rewritten in 
the conventional Landau-Ginzburg form (in the London limit). 
Moreover if one chooses the bare CS coefficient to be
unit, leaving the CS matter coupling constant $e$ 
(the statistical charge) free, 
and denotes the vacuum polarization associated with the CS term by
$\Pi_o=\Pi_o(e,p^2=0)$, the criterion takes the form
\begin{equation}
                        \tilde{\Pi}_o = 1 + \Pi_o = 0 
\label{cri}
\end{equation}
Namely, the matter-induced CS term cancels out the bare one  exactly. 
Eq.(\ref{cri}) determines 
the critical statistical charge, $e_c$, of the superconducting phase
of the system.
 
In this paper we examine the criterion in certain field theory models. 
In particular, we will discuss
a method of using the Dyson-Schwinger (DS) equations to 
calculate the critical statistical charge, $e_c$, for massless
theories. To be concrete, we will use the CS massless fermion theory 
as an example. A parallel discussion for the CS massless boson theory
is straightforward. 

We start with a brief review of relevant perturbative results and point
out the difference between the cases of the massive and massless theories.
Let us consider the $U(1)$ gauge theory with a CS field minimally coupled 
to a fermion field and a boson field in the three dimensional Euclidean 
space ($g_{\mu\nu}=\delta_{\mu\nu}$):
\begin{equation}
{\cal L} =  \psi^\dagger
[\gamma_\mu(\partial_\mu - iea_\mu) + iM]\psi 
+(\partial_\mu + iea_\mu)\phi^*(\partial_\mu - iea_\mu)\phi + m^2\phi^*\phi
- 
i\epsilon_{\mu\nu\lambda}a_\mu\partial_\nu a_\lambda\;
\label{L}
\end{equation}
where the two-by-two gamma matrices $\gamma_\mu$ 
are anti-hermitean, $\gamma_\mu^\dagger 
= -\gamma_\mu$; $\psi$ is a two-component spinor, $\phi$ is a complex scalar,
and $a_\mu$ is the Chern-Simons gauge field. $M$ and $m$
are the fermion and boson mass, respectively. The coupling constant $e$
is dimensionless by naive powercounting (For simplicity we use $e$ to 
denote the couplings for both the 
CS scalar and CS spinor, knowing the two interactions are not necessarily
to have the same strength).
The symmetries of the theory have been discussed 
in \cite{DJT}.  
In particular, the charge conjugation (C) transformation 
leaves the lagrangian
invariant, while the fermion mass and the Chern-Simons terms are both
variant under parity (P) or time reversal (T). But CPT symmetry 
holds.

For the massive matter fields ($M$ and $m$ are non-zero), there is
a no-renormalization theorem \cite{CH}\cite{SSW}:  
there is no radiative correction to the CS term beyond one-loop. 
Consequently, one needs to compute only one-loop diagrams to
determine $\Pi_o$. The one-loop results are
\cite{NS} 
\begin{eqnarray}
         \Pi_o &=& \Pi_o^{l=1} \equiv 0, 
~~~{\rm from~~a~~massive~~boson} \label{massive1}\\ 
         \Pi_o &=& \Pi_o^{l=1}(p^2=0) 
       = -\frac{e^2}{4\pi}{\rm sign}(M), ~~~{\rm from~~a~~massive~~ 
fermion} \label{massive2}
\end{eqnarray}
From the above results, one may draw the conclusion that the massive 
fermions in the minimal coupling induce a CS term of the 
proper sign and inevitably exhibit superconductivity 
at some critical value of the statistical charge, while 
the massive bosons do not. 

On the other hand, if the coupled matters are massless ($M = m =0$), 
the no-renormalization theorem is 
invalid \cite{SSW}. Explicit computation shows that massless matters,
either massless bosons or massless fermions, do not contribute to 
the CS term at one loop,  
but make finitely contributions to it at two-loop  \cite{Che}:
\begin{eqnarray}
                \Pi_o^{l=2} &=& -\frac{e^4}{16}
(\frac{1}{4}-\frac{1}{\pi^2})~~~~ {\rm from~~a ~~massless~~ boson} 
\label{massless1}\\ 
                \Pi_o^{l=2} &=& -\frac{e^4}{4}
(\frac{1}{16}+\frac{1}{\pi^2})~~~~{\rm from~~a ~~massless~~fermion}  
\label{massless2}
\end{eqnarray}
Notice that, in the massless cases, $\Pi_o$ is independent of the external 
momentum $p$. Four loop and higher order corrections are expected. 
We see now that the results from the massless matters
are 
even qualitatively different 
from those of the massive ones \cite{CFW}. 
In particular, a CS massless boson theory obviously 
exhibits the anyonic superconductivity too. 

Interesting enough, the CS massless matter
field theories  at the critical points 
may be regarded as models of  the P and T conserved superconductivity. 
Indeed, in the superconducting phase, the renormalized CS term - 
the only term that violates P and T - 
vanishes and the P and T symmetries are recovered. 
Saying so, we have actually assumed that the massless matter fields, 
especially fermion fields, in the CS quantum theory remain massless, 
a point we will further address later. 

From the two-loop results, Eqs.(\ref{massless1}) and (\ref{massless2}), 
it is attempting to calculate the critical values of the statistical charge.
These are $e_c =\pm 2.223$ for a fermion and $e_c=\pm 3.221$ for  
a boson. The two-loop results of such strong critical couplings 
raise concern on whether a perturbation expansion in $e$ is applicable  
for studying the physics near and in the superconducting phase. 
Therefore, in this case, non-perturbative methods must be used. The
one we will use is to solve the DS equations.

The DS equations, in principle, 
contain all information about the quantum field theory studied,
and provide a natural non-perturbative scheme to study its dynamics. 
However, since the DS equations are a set of infinitely many coupled 
equations, one needs to make certain truncation to get 
a tractable subset. Some features of 
(2+1) dimensional QED without or with a CS term have been
studied by using appropriate approximations in the DS equation for
the fermion propagator \cite{App}-\cite{KK}. 
To the problem we are considering, it is necessary (and sufficient) 
to consider the DS equations for the two-point Green functions 
 $\Delta_{\mu\nu}(p)$ for the CS field and 
$iS(p)$ for the fermion field 
(from now on, we focus on the CS massless fermion theory). 
Denoting the inverse of the two two-point Green functions by 
$\Pi_{\mu\nu}(p)$ and $iS^{-1} (p)$, 
respectively, 
we have the two DS Equations
\begin{eqnarray}
               \Pi_{\mu\nu} (p) &=& \Pi_{\mu\nu}^0(p) 
      - e^2\int \frac{d^3k}{(2\pi)^3}\gamma_\mu S(k+p)\Gamma_\nu (k+p,k)S(k)
\label{DS1}\\
                     S^{-1} (p) &=& S_0^{-1}(p) -e^2\int \frac{d^3k}{(2\pi)^3}
              \gamma_\mu S(k+p)\Gamma_\nu (k+p,p)\Delta_{\nu\mu}(k)
\label{DS2}\end{eqnarray}
where $\Gamma_\mu(p,q)$ is the full fermion-photon-fermion 
three-point function (the vertex function), which we will discuss 
in details later; and the bare quantities (in the Landau gauge) are
\begin{eqnarray}
              \Pi^0_{\mu\nu}(p) &=& \epsilon_{\mu\nu\sigma}p_\sigma,~~~~~~~~~
~~              iS^{-1}_0 (p) = ip\cdot \gamma\\
     \Delta^0_{\mu\nu} (p) &=& - \frac{\epsilon_{\mu\nu\sigma}p_\sigma}{p^2},
~~~~~~~~~~ iS_0(p) =i\frac{p\cdot \gamma}{p^2}
\label{DS}\end{eqnarray}

At this point, we would like to make several remarks. 
First of all, with only a dimensionless 
coupling constant $e$, the CS fermion theory is renormalizable. 
We will use the regularization by dimensional reduction,  
in which one performs a dimensional continuation of the integration 
measure in all Feynman integrals, but the vector and the tensor 
quantities including the Levi-Civita tensor $\epsilon_{\mu\nu\sigma}$ 
are always treated as if they are strictly three dimensional 
\cite{Che}\cite{CSW}. 
Secondly, 
we have chosen the Landau gauge for the gauge fixing. It is known that
in this gauge choice the CS matter theory is explicitly free of infrared 
divergence \cite{PR}. Indeed, in the Landau gauge, 
the CS propagator takes the form of $p^{-1}$ as $p \rightarrow 0$, as shown
in Eq.(\ref{DS}), 
contrast to  $p^{-2}$  for the propagator of  the gauge field in the 
conventional QED$_3$. 
Third, it has been found that 
the $\beta$-function for the CS coupling vanishes up to three loops 
in various kinds of CS matter theories  \cite{Che}\cite{CSW}. 
Since the kinetic CS action is of topological nature, 
it is reasonable to conjecture that the $\beta$-function vanishes  
to all orders. For a massless CS theory, who's lagrangian is scale 
and (global) conformal invariant, the vanishing $\beta-$function 
implies that the scale of conformal 
symmetries survive quantization and renormalization \cite{CSW}. 
Consequently, all physical quantities are  
independent of the renormalization scale, 
$\mu$  (with dimension of mass), which is routinely 
introduced with the regularization scheme. 
In particular, 
the matter field and the CS field  
receive no radiative mass corrections. 
To see this, let $m(e,\mu)$ be the renormalized mass of, say, 
the matter field. By dimensional argument, 
the renormalized mass should be $\mu$ times a function of $e$ \cite{Col}. 
Since any physical quantity is invariant under renormalization 
group transformation, we have
\begin{equation}
     0=\mu\frac{d}{d\mu}m(e,\mu) = m + \beta\frac{\partial}{\partial e}m
\end{equation} 
It shows that $m = 0$ when $\beta=0$. 
Finally, we list the one loop corrections to the polarization tensor of the 
``photon'', the  fermion self-energy and the vertex function:
\begin{eqnarray} 
      \Pi^{l=1}_{\mu\nu}(p) &=& -\frac{e^2}{16p}P_{\mu\nu},~~~~~~~
      i\Sigma^{l=1}(p) = -i\frac{e^2}{8}p \label{one1}\\
      i\Gamma^{l=1}_\nu(p,q) &=&i\frac{e^2}{8}[\frac{p_\nu+q_\nu}{p+q}
                               +\Gamma^{t(l=1)}_\nu(p,q)] \label{one2}\\
 \Gamma^{t(l=1)}_\nu(p,q)&=&-2\frac{p_\mu(q^2-p\cdot q)+q_\nu(p^2-p\cdot k)}
                     {(p+q)|p-q|(p+q+|p-q|)}
  -\epsilon_{\nu\sigma\eta}(q-p)_\sigma \gamma_\eta \frac{1}{p+q+|p-q|}
\nonumber
\end{eqnarray}
\begin{equation}
+[(pq+p\cdot q)\epsilon_{\nu\sigma\eta}
(\frac{q_\sigma}{q}-\frac{p_\sigma}{p}) 
+(\frac{q_\nu}{q}+\frac{p_\nu}{p})\epsilon_{\sigma\tau\eta}q_\sigma p_\tau
]\gamma_\eta\frac{1}{(p+q+|p-q|)^2} 
\label{one3}\end{equation}
where $P_{\mu\nu} = \delta_{\mu\nu}p^2-p_\mu p_\nu$ and $p = |p|$. 
These in Eq.(\ref{one1}) are well known, 
while Eqs.(\ref{one2}) and (\ref{one3}) are new results.

Now we consider the general forms of $\Pi_{\mu\nu}(p)$ and 
$S^{-1}(p)$.  By the dimension argument, the gauge symmetry and 
the (global) conformal symmetry, we find that  
the inverse two-point functions take the forms
\begin{eqnarray}
               \Pi_{\mu\nu} (p) &=& 
\Pi_eP_{\mu\nu}/p+\tilde{\Pi}_o\epsilon_{\mu\nu\sigma}p_\sigma\label{Pi}
\\
            S^{-1}(p)& =& Ap\cdot\gamma + Bp
\label{S-}
\end{eqnarray}
where $\Pi_e$, $\tilde{\Pi}_o$, $A$ and $B$ are 
independent, dimensionless, and finite constants.
Accordingly, we have 
\begin{eqnarray}
                \Delta_{\mu\nu} (p) &=& \frac{1}{\Pi_e (1+\theta^2) p^3}
 (P_{\mu\nu}-\theta p\tilde{\Pi}_o\epsilon_{\mu\nu\sigma}p_\sigma)
\label{Del}\\
           S(p)& =& \frac{1}{A^2+B^2}\frac{Ap\cdot\gamma - Bp}{p^2}
\label{Ep}
\end{eqnarray}
with $\theta=\tilde{\Pi}_o/\Pi_e$. 
Eqs.(\ref{Del}) and (\ref{Ep}), involve no more singularity 
except the pole at $p^2=0$, which is consistent with the argument given above
that the matter and photon remain massless.

Each of the DS equations (\ref{DS1}) and (\ref{DS2}) 
consists of two independent equations due to the two-by-two $\gamma$-matrices. 
Contracting Eq.(\ref{DS1}) with $P_{\mu\nu}(p)$ 
and with $\epsilon_{\mu\nu\tau}p_\tau$, respectively, we obtain
\begin{eqnarray}
              2p^3 \Pi_e& =&
    - e^2P_{\mu\nu}(p) \int \frac{d^3k}{(2\pi)^3}
   \gamma_\mu S(k+p)\Gamma_\nu (k+p,k)S(k) \label{a1}\\
              2p^2 (\tilde{\Pi}_o-1)& =&
    -e^2\epsilon_{\mu\nu\tau}p_\tau \int \frac{d^3k}{(2\pi)^3}
   \gamma_\mu S(k+p)\Gamma_\nu (k+p,k)S(k)\label{a2}
\end{eqnarray}  
Similarly, taking the trace of Eq.(\ref{DS2}) and  the trace of
Eq.(\ref{DS2}) multiplied by $p\cdot\gamma$, 
we obtain
\begin{eqnarray}
                     2pB &=& -e^2Tr\int \frac{d^3k}{(2\pi)^3}
 \gamma_\mu S(k+p)\Gamma_\nu (k+p,p)\Delta_{\nu\mu}(k)\label{a3}\\
                     2p^2(A-1) &=& e^2Tr\int \frac{d^3k}{(2\pi)^3}
         p\cdot \gamma \gamma_\mu S(k+p)\Gamma_\nu (k+p,p)
\Delta_{\nu\mu}(k)\label{a4}
\end{eqnarray}

Our problem is to solve the constants A, B, $\Pi_o$ and $\Pi_e$ from 
the set of Eqs.(\ref{a1}-\ref{a4}). For this purpose, 
we need to know the three-point function, $\Gamma_\nu (p,q)$. 
$\Gamma_\nu (p,q)$ receives the decomposition:
\begin{equation}
    \Gamma_\nu(p,q) = \Gamma^L_\nu(p,q)+\Gamma^T_\nu(p,q)
\label{ver}
\end{equation}
The longitudinal part, $\Gamma^L_\nu(p,q)$, is completely 
determined by the inverse fermion two-point function 
via the well-known Ward-Takahashi identity 
for the $U(1)$ symmetry
\begin{equation}
              (p-q)_\nu\Gamma^L_\nu(p,q) = S^{-1}(p) - S^{-1}(q)
\label{WT}
\end{equation}
Using Eq.(\ref{S-}), we have  
\begin{equation}
      \Gamma^L_\nu(p,q) = A\gamma_\nu + B \frac{p_\nu+q_\nu}{p+q}
\label{Log}
\end{equation}
But the transverse part of the vertex function, $\Gamma^T_\nu(p,q)$, 
satisfying
\begin{equation}
(q-p)_\nu\Gamma^T_\nu(p,q) = 0
\label{Tra}
\end{equation}
is less constrained. The simplest ansatz $\Gamma^T_\nu=0$ does not
work as the pure longitudinal vertex
Eq.(\ref{Log}) has been found to introduce singularities into  
Eqs.(\ref{a1}-\ref{a4}).
As a matter of fact, even at one loop level of the perturbation theory, 
from Eq.(\ref{one2}) we have seen $\Gamma^T_\nu\neq0$.

In \cite{BC} and \cite{CP}, the form of the tensor structure 
of the transverse part of the vertex in (3+1) dimensional QED was discussed,
and eight linearly independent terms for the transverse vertex 
were given. The analysis in \cite{BC} and \cite{CP} 
applies to the 2+1 dimensional CS gauge theory as well. In particular, 
one of the eight transverse vectors, 
$T_8 =-\gamma_\nu p_\eta q_\tau \sigma_{\eta\tau}
+p_\nu q\cdot \gamma - q_\nu p\cdot\gamma$ 
with $\sigma_{\eta\tau} = \frac{1}{2}[\gamma_\eta, \gamma_\tau]$,
turns out to be a pure axial vector. 
In (2+1) dimensions, it can be written
$T_8=W_\nu=\epsilon_{\nu \sigma \eta}p_\sigma q_\eta$. 
We would require the vertex function to be 1)  
gauge invariant so that Eq.(\ref{Tra}) holds; 2) 
the charge conjugation invariant such that it satisfies
(the charge conjugation operator $C = \gamma_2$):
\begin{equation}
                 C^{-1}\Gamma_\nu (p,q) C = [\Gamma_\nu (-q,-p)]^t
\end{equation}
and 3) free of kinematic singularities \cite{BC}. 
Then the ansatz we will make for the transverse vertex is: 
\begin{eqnarray}
& &       \Gamma_\nu^T(p,q)
= BQ_\nu\frac{2}{(p+q)|p-q|(p+q+|p-q|)} + C\Gamma_{\nu\sigma}^\epsilon (p,q)
\gamma_\sigma\nonumber\\ 
& & + DQ_\nu\frac{(p+q)\cdot\gamma}{pq|p-q|^2}
+EQ_\nu \frac{W \cdot \gamma}
 {pq|p-q|^3}+F W_\nu\frac{1}{|p-q|(p+q+|p-q|)} 
\label{Gamma}
\end{eqnarray}
where
\begin{eqnarray}
        Q_\nu(p,q)&=&p_\nu q\cdot(p-q)-q_\nu p\cdot (p-q),~~~~~	 
        W_\nu (p,q)= \epsilon_{\nu \eta \tau}p_\eta q_\tau\\ 
       \Gamma_{\nu\sigma}^\epsilon (p,q)
&=&[(pq+p\cdot q) \epsilon_{\nu\eta\sigma}
    (\frac{q_\eta}{q}-\frac{p_\eta}{p})
    -(\frac{q_\nu}{q}+\frac{p_\nu}{p})W_\sigma]\frac{1}{pq}     
\end{eqnarray}
Note that 
the forms of the first two terms in Eq.(\ref{Gamma}) have been seen 
in the one loop correction Eq.(\ref{one3}) and  
the other terms might come from higher loop corrections,
though we are seeking a non-perturbative solution.
The constants C, D, E and F will be fixed by demanding
that the ultraviolet divergence in the set of Eqs.(\ref{a1}-\ref{a4})
cancel out.

With the ansatz for the full vertex function given in 
Eqs.(\ref{ver}), (\ref{Log}) and (\ref{Gamma}),
the set of equations (\ref{a1}-\ref{a4}) 
for the variables $A$, $B$, $\Pi_e$, and 
$\tilde{\Pi}_o$ are highly non-linear and inhomogeneous.
The general solutions of the set of equations 
will be discussed elsewhere. Here, to obtain 
the critical statistical charge for the superconduct phase,
 we use the condition $\tilde{\Pi}_o=0$ (and then $\theta = 0$). 

Substituting Eqs.(\ref{Del}), (\ref{Ep}), (\ref{Log}) 
and (\ref{Gamma}) in Eqs.(\ref{a1}-\ref{a4}) 
and performing the integrations over k with the regularization 
by dimensional reduction, we obtain a set of algebra equations
(with $\tilde{\Pi}_o=0$): 
\begin{eqnarray}
         \Pi_e&=& \frac{e^2}{16(A^2+B^2)^2}[-A^3-\frac{4}{\pi^2}AB^2 
-\frac{8}{\pi^2}A^2D+B^2D-(\frac{4}{\pi^2}-1)A^2F])\label{Pie}\\
      1 &=& \frac{e^2}{32(A^2+B^2)^2}
 [(2-\frac{8}{\pi^2})A^2B +\frac{16}{\pi^2}A^2E +B^2E+\frac{8}{\pi^2}ABF]
\label{Pie2}
\\
    B&=& \frac{Ae^2}{48\Pi_e(A^2+B^2)} 
   [(3-\frac{12}{\pi^2})B - \frac{12}{\pi^2}C +\frac{8}{\pi^2}E]
\label{Pie3}
\end{eqnarray}
\begin{equation}
    A=1+\frac{e^2}{64\Pi_e (A^2+B^2)} 
   [-\frac{32}{\pi^2}B^2+\frac{48}{\pi^2}BC-\frac{16}{\pi^2}AD
         +BE+(\frac{16}{\pi^2}-4)AF]
\label{Pie4}
\end{equation}
with the finiteness conditions 
\begin{eqnarray}
          0 &=& 5(B^2-AF)+2BE \label{fc1}\\
          0 &=& A^2B-2(A^2+B^2)C - 2ABD +ABF \label{fc2}\\
          0 &=& 2BD - AE \label{fc3}\\
          0 &=& 2A^2+B^2+2AF
\label{fc4}
\end{eqnarray}
Using the conditions (\ref{fc1}-\ref{fc4}), 
eliminating $\Pi_e$ by Eq.(\ref{Pie}), and setting $A=xB$
in the Eqs. (\ref{Pie2}-\ref{Pie4}), we obtain
\begin{eqnarray}
      0 &=& 520x^4+(1132-102\pi^2)x^2+264 - 81\pi^2\\
      0 &=& [80x^6-(58\pi ^2+72)x^4-(37\pi^2+288)x^2+7\pi^2+208]B\nonumber\\
& &+2x(-80x^4+(26\pi^2-152)x^2+19\pi^2+16)\\
      0 &=& [160x^4+(2\pi ^2+304)x^2+15\pi ^2+16]e^2
          +128\pi ^2(x^2+1)^2B
\end{eqnarray}
The unique physical solution is (in the other solutions, either $x^2$ 
or $e_c^2$ has a wrong sign)
\begin{equation}
                    e_c =\pm 2.106 
\end{equation}    
companying with $A=0.5404$, $B=-0.5692$ and $\Pi_e=-0.3773$. 

It is interesting to notice that the critical statistical charge
obtained here with the non-perturbative method differs from 
the two-loop result by merely $5.6\%$. 

This work was supported in part by U.S. NSF and by U.S. DOE.

\end{document}